\documentclass[conference,letterpaper]{IEEEtran}
\IEEEoverridecommandlockouts
\usepackage{cite}
\usepackage{amsmath,amssymb,amsfonts}
\usepackage{algorithmic}
\usepackage{graphicx}
\usepackage{textcomp}
\usepackage{physics}
\usepackage{amsthm}
\usepackage{xcolor}
\usepackage{enumitem}
\usepackage{subcaption}
\usepackage{balance}
\def\BibTeX{{\operatorname B\kern-.05em{\sc i\kern-.025em b}\kern-.08em
    T\kern-.1667em\lower.7ex\hbox{E}\kern-.125emX}}
\begin{document}

\title{Proving Erasure\\
\thanks{This work is supported by the Swiss National Science Foundation (SNF)  and the Fonds de recherche du Qu\'ebec -- Nature et technologies (FRQNT).}}

\newtheorem{definition}{Definition}
\newtheorem{proposition}{Proposition}
\newtheorem{theorem}{Theorem}
\newtheorem{lemma}{Lemma}
\newcommand{\VPA}{(\mathcal{V},\mathcal{P},\mathcal{A})}

\author{\IEEEauthorblockN{Xavier Coiteux-Roy}
\IEEEauthorblockA{\textit{Faculty of Informatics} \\
\textit{Universit\`a della Svizzera italiana}\\
Lugano, Switzerland \\
xavier.coiteux.roy@usi.ch}
\and
\IEEEauthorblockN{Stefan Wolf}
\IEEEauthorblockA{\textit{Faculty of Informatics} \\
\textit{Universit\`a della Svizzera italiana}\\
Lugano, Switzerland \\
stefan.wolf@usi.ch}
}

\maketitle
\begin{abstract}
It seems impossible to certify that a remote hosting service does not leak its users' data{~---~}or does quantum mechanics make it possible? We investigate if a server hosting data can information-theoretically prove its definite deletion using a ``BB84-like'' protocol.
To do so, we first rigorously introduce an alternative to privacy by encryption: {\em privacy delegation}.
We then apply this novel concept to provable deletion and remote data storage. For both tasks, we present a protocol, sketch its partial security, and display its vulnerability to eavesdropping attacks targeting only a few bits.

\end{abstract}


\section{Introduction}
As the story goes~\cite{ashley93}, some sailors who did not have access to secure private storage on ships would tie their bag of belongings with the ``thief knot.'' This particular knot resembled the reef knot, one of the most common knots, so that anyone not knowing the secret would tie back the bag the wrong way. The canny sailors would then detect when someone went through their belongings because the thief knot they tied would be altered.

This folktale inspires us to introduce in Section~\ref{privacy} a novel, relaxed, notion of privacy: \emph{privacy delegation}. Privacy delegation does not prevent eavesdropping but it makes such an act inevitably detectable. It is useful in applications where we cannot aspire to have perfect encryption in the sense of Shannon\cite{shannon}.

A typical application is data storage (Section~\ref{storage}): Information-theoretically secure encryption of data is impossible unless one keeps a secret key at least as long as the message that is remotely stored\cite{shannon}, defeating the purpose of storing it in the first place. So how can a hosting server meaningfully certify it protected one's privacy?

This is related to the question of provable erasure (Section~\ref{deletion}): Can one verify some information was indeed deleted? These two tasks seem at first sight impossible, at least when restricting ourselves to classical physics. We use quantum theory instead and follow the path of Bennett and Brassard in 1984\cite{BB84}. We present a straightforward protocol in Section~\ref{naive} and sketch a proof of its partial security in Section~\ref{partialsec}, before showing in Section~\ref{wattack} it fails to be unconditionally secure against some limited attacks. We are, in Section~\ref{privacyampl}, left with the question of how to fix it.

\section{Delegating Privacy}\label{privacy}

\subsection{Privacy Delegation: an Alternative Standard for Privacy}
\begin{figure}[htbp]
\begin{subfigure}{0.5\textwidth}
\includegraphics[width=\linewidth]{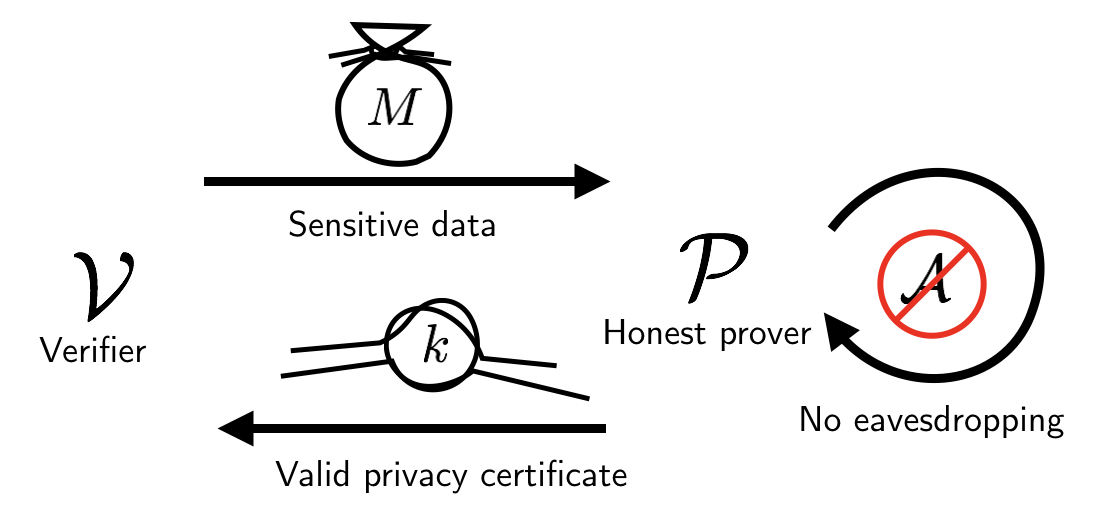}
\caption{A valid privacy certificate guarantees that no leaks about the sensitive message occured.} \label{fig:1a}
\end{subfigure}
\hspace*{\fill} 
\begin{subfigure}{0.5\textwidth}
\includegraphics[width=\linewidth]{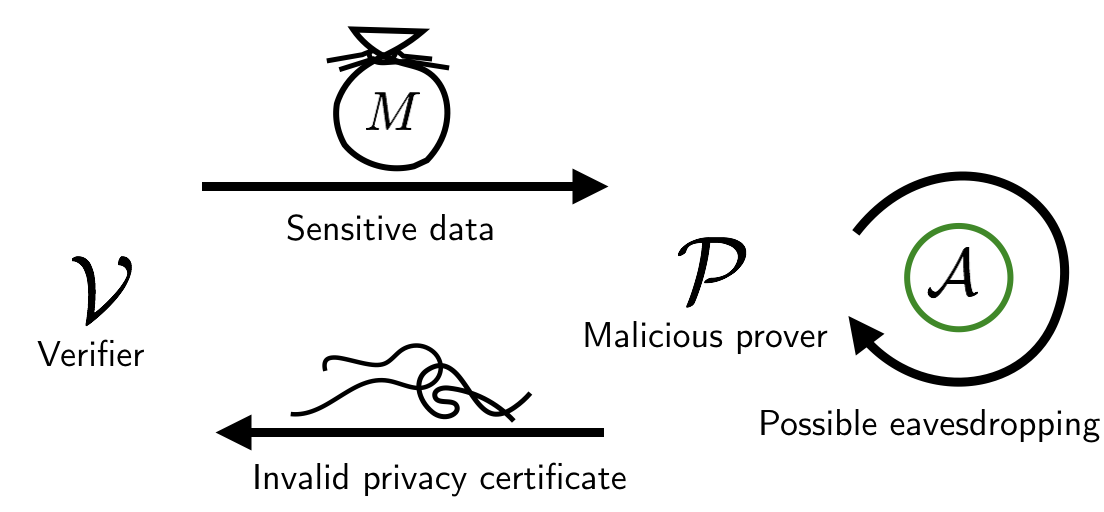}
\caption{It is arbritrarily unlikely to produce a valid privacy certificate if eavesdropping has occured.} \label{fig:1b}
\end{subfigure}
\caption{Privacy delegation is a novel approach to privacy.} \label{fig:1}
\end{figure}

The general framework of privacy delegation is the following. After a certain protocol ({\em e.g.}, data storage), the prover/server presents a proof to the verifier/user.
If the proof is accepted by the verifier, they are sure the prover respected their privacy and that none of their data was leaked.
On the other hand, the rejection of the proof by the verifier alarms them that their data might be compromised. Fig.~\ref{fig:1} summarizes the idea of delegating privacy.

\begin{definition}
A privacy-delegation protocol is an interactive protocol between a prover $\mathcal{P}$ and a verifier $\mathcal{V}$ that aims to establish whether a message $M$, sent by $\mathcal{V}$ to be temporarily held by $\mathcal{P}$, was rigorously protected from any past or future eavesdropper~$\mathcal{A}$. $\mathcal{V}$ takes as input the message $M$ and the security parameter $n$. $\VPA$ are modelled formally as probabilistic Turing machines.
\end{definition}

We will define secure privacy delegation in the language of modern cryptography\cite{lindell2014introduction} through the notion of conditional indistinguishability: We will say the protocol is secure if, once the privacy certificate has been produced by the prover~$\mathcal{P}$ and accepted by the verifier~$\mathcal{V}$, no eavesdropping by an adversary~$\mathcal{A}$ can offer any advantage in discriminating whether the now-completed privacy-delegation protocol was run on one message, or another, given the two messages are of the same length.
\begin{definition}
A privacy-delegation protocol is information-theoretically secure if and only if, for every prover $\mathcal{P}$ and adversary $\mathcal{A}$,
\begin{multline}  
\operatorname{P}\!\left[\operatorname{CERT}^{\VPA}\!(n)\right] \cdot \left( \operatorname{P}\!\left[\operatorname{DISCR}^{\VPA}(n)|\operatorname{CERT}^{\VPA}\!(n)\right]\right.\\-1/2\Big)< \operatorname{negl}(n)\,,\tag{security}
\end{multline}
\\
where $n$ is the security parameter and $\operatorname{negl}(n)$ is a negligible function, meaning smaller than the inverse of any polynomial function of $n$ for sufficiently large $n$.
\end{definition}
$\operatorname{CERT}^{\VPA}$ is the certification experiment and is passed ({\em i.e.}, it evaluates to $\operatorname{true}$) if and only if the verifier $\mathcal{V}$ accepts the $\emph{privacy certificate}$ produced by the prover $\mathcal{P}$.

$\operatorname{DISCR}^{\VPA}$ is the discrimination experiment, and it is passed if and only if, when attacking the privacy-delegation protocol executed on either the legitimate message or a dummy one (with equal probability), the adversary $\mathcal{A}$ guesses successfully which one it is. While not necessary, it is convenient to pose~$\mathcal{P}$ and~$\mathcal{A}$ as the same entity.

Note that conditioning $\operatorname{DISCR}^{\VPA}$ on $\operatorname{CERT}^{\VPA}$ does not imply that the certification experiment is necessarily done first, as a valid privacy certificate implies eavesdropping can neither have occurred before, nor can occur after.

Finally, the compact formulation of the privacy-delegation security definition reflects that the privacy-delegation protocol is secure as soon as either term is negligible.

Note how, in presence of a prover generating a valid privacy certificate with probability $1$, the above definition of secure privacy delegation reduces to Shannon's notion of information-theoretic security against eavesdropping. Any scheme which is perfectly secure in Shannon's sense can be seen as a trivial privacy-delegation scheme for which all certificates are valid.

We conclude our definition of privacy delegation by requiring one additional property: It is desirable that a privacy-delegation protocol fails with non-negligible probability only when the prover is dishonest.
\begin{definition}
A privacy-delegation protocol is correct if and only if there exists a prover ${\mathcal{P}_{\text{honest}}}$ such that
\begin{equation}
\operatorname{P}\!\left[\operatorname{CERT}^{(\mathcal{V},\mathcal{P}_{\text{honest}},\mathcal{A})}\!(n)\right]>1-\operatorname{negl}(n) \,. \tag{correctness}
\end{equation}
\end{definition}
\subsection{Privacy Delegation Is an Alarm Bell}
We shall stress the role of privacy delegation to detect information leaks; it is not by itself an encryption method: It offers no security against a malicious prover other than announcing its behaviour. In this sense, privacy delegation is more an alarm bell than a lock.

It can be useful in bringing a concept of liability to a hosting server. When a server leaks its users' data, whether by malice or negligence, its privacy certificate will be rejected, and the server can subsequently be boycotted or taken to court.

Note that privacy delegation is different from digital watermarking\cite{watermarking}, even if both techniques can be used to detect leaking agents. Digital watermarking mixes the sensitive data with hidden identification information in a way the source of the leak can be identified once the leaked data surfaces, while privacy delegation enables to detect misbehaviour even if no abnormal traces of the sensitive information have been observed. Privacy delegation offers in that sense stronger protection, even if neither approach directly prevents the leakage of data.

\section{Applications}
\subsection{Remote Storage}\label{storage}
An important application of privacy delegation is the remote storage of classical information. In this setting, a user wants to upload classical data to a server and be able to retrieve it later; while the server wants to prove to the user it protected its data from any eavesdropping ---~{\em both in the past and in the future}. We explicit the general form of this task.\\

\noindent \textbf{The task of remote storage with privacy delegation:}
\begin{enumerate}[nolistsep]
\item A user (verifier $\mathcal{V}$) wants to store the classical message~$M$ on a server (prover~$\mathcal{P}$). The user can manipulate quantum states.
\item $\mathcal{V}$ generates a key~$k$ at random and sends quantum encoding $\rho(M,k)$ to~$\mathcal{P}$. The state $\rho(M,k)$ depends on the specific protocol.
\item Time passes. Eventually the user~$\mathcal{V}$ asks the server~$\mathcal{P}$ for its data~$\rho(M,k)$ back and wants to be assured no copies were made.
\item $\mathcal{P}$ sends back a quantum state $\rho'$ which will also act as the privacy certificate. If they are honest, $\rho'=\rho(M,k)$.
\item $\mathcal{V}$ examines the privacy certificate ({\em i.e.}, $\rho'$). If they accept it and the protocol is secure, neither $\mathcal{P}$ nor any eavesdropper $\mathcal{A}$ have any information about $M$.\\
\end{enumerate}

Note that the last step is not a full authentication of the quantum state $\rho(M,k)$ but only of its privacy-delegation layer\footnote{We were made aware after finishing this work of Daniel Gottesman's concept of \emph{uncloneable encryption}, and of its conjecture that it could be applicable to schemes that were different from quantum authentication schemes in that they would ``not authenticate the classical message''~\cite{gottesman2002uncloneable}. Since our privacy-delegation scheme does no error estimation in the rectilinear basis, our work is coincidentally relevant to that interrogation. We do not directly study the impact of revealing the user's key at the end of the privacy-delegation protocols, but it would be a very relevant question for this framework---it would make erasure even more meaningful.}. It is, however, easy to guarantee the integrity of the classical data by adding on top of the privacy-delegation protocol a Wegman-Carter classical authentication \mbox{scheme\cite{wegman1981new}.} These schemes are well-known and can be information-theoretically secure even with a short key.

Because of the nature of remote storage with privacy, it is clear that if the user must keep a secret key: It must at least be shorter than the message to be stored.
This is impossible if we aim for arbitrarily perfect secrecy, as stated by Shannon's theorem, and it is why we look for privacy delegation instead of encryption.

Privacy delegation differs from, but is not incompatible with (as we develop next), the standard approach which is to aim for computational security. Computational security is reached through classical encryption relying on assumptions on the computing power of the adversary and the hardness of certain mathematical problems\cite{aes,3des}.

\subsection{Provable Deletion}\label{deletion}

Provable deletion can be framed as a general case of remote storage. In this scenario, the user/verifier is not interested in retrieving their remotely stored data, but simply orders for its definitive erasure. The server/prover subsequently produces a token which, if accepted by the user, certifies the erasure.
It is of special interest because classically, there is no way for the server to prove it did not secretly keep a backup of the data, as the act of copying classical information is undetectable.\\

\noindent \textbf{The task of provable deletion with privacy delegation:}
\begin{enumerate}[nolistsep]
\item A user (verifier~$\mathcal{V}$) wants to store the classical message~$M$ on a server (prover~$\mathcal{P}$). The user can manipulate quantum states. 
\item $\mathcal{V}$ generates a key $k$ at random and sends quantum encoding $\rho(M,k)$ to~$\mathcal{P}$. The state $\rho(M,k)$ depends on the specific protocol.
\item Time passes. Eventually the user $\mathcal{V}$ asks the server~$\mathcal{P}$ to provably delete all of its data~$M$.
\item $\mathcal{P}$ applies some operations on~$\rho(M,k)$ and provides to~$\mathcal{V}$ a privacy certificate~$C$.
\item $\mathcal{V}$ examines $C$, whose acceptance proves the erasure of~$M$.\\
\end{enumerate}

We note here that we are analyzing the provable deletion of purely classical information. The no-go theorem for deleting \emph{arbitrary} quantum states\cite{pati2000impossibility} does, therefore, not apply.

\subsection{Combining Privacy Delegation and Computational Security}
In cases where standard privacy is desirable, meaning when privacy concerns require more than leak detection, privacy delegation can be complemented by classical encryption with computational security. Such a combination has multiple advantages.
First, the storage is still meaningful as both techniques only necessitate short keys.
Second, it makes possible the concept of recalling encrypted data. Standard classical encryption relies on assumptions that need not hold forever: Computing power grows exponentially, and the underlying hard mathematical problems could at any time be solved more easily than currently believed. There is no guarantee of everlasting security with standard classical encryption because any encrypted message can be stored by an adversary until its encryption becomes obsolete\cite{rabin}. As such, there is always a risk, even if one uses strong up-to-date encryption, in uploading data to a server that could leak it. If, however, the server is able to produce a valid privacy certificate when the user wants to recall data to change its encryption, then the user can be sure that her or his data is still perfectly safe. 

Finally, an adversary scanning massive amounts of data looking for weak encryptions, obsolete standards and/or valuable information can be quickly detected by the privacy-delegation scheme. This makes the hosting network more secure as a whole, since an adversary will not necessarily know what type of computational encryption they are attacking before they attack it in a detectable way.

\section{Implementation: A Naive Protocol}\label{naive}
We explore how to implement privacy delegation in the information-theoretically secure way of Section~\ref{privacy}. We start with a naive protocol which splits into two versions: remote storage ($\operatorname{STORAGE}$) and provable deletion ($\operatorname{ERASURE}$). We then show how they are partially secure whereas an eavesdropper can still gain {\em some\/} small amount of information even when privacy should be certified.

\subsection{The Encoding}
\begin{figure}[htbp]
\includegraphics[width=\linewidth]{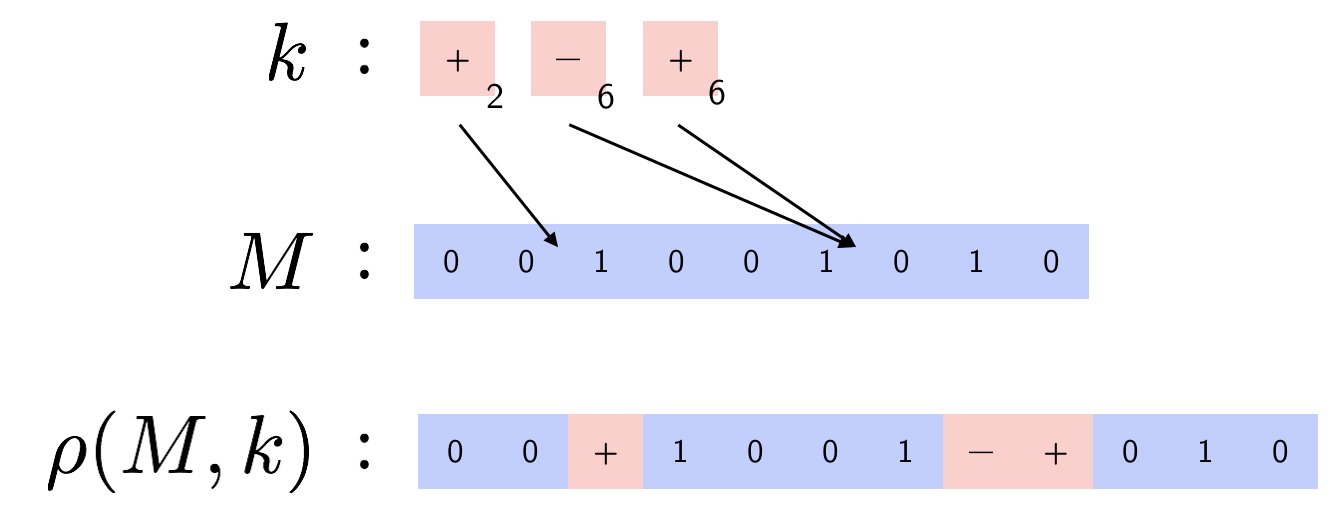}
\caption{The message is in the rectilinear basis and is interceded with bits in the diagonal basis whose ulterior integrity certifies the privacy of the then-recovered (or then-erased) message.} \label{fig:2}
\end{figure}

Both versions start the same way. The idea is for the user to pre\"emptively, at random positions, sprinkle the $m$ bits plain message in the rectilinear basis $\{\ket{0},\ket{1}\}$ with $n$ random ``trap bits'' in the diagonal basis $\{\ket{+},\ket{-}\}$. Fig.~\ref{fig:2} illustrates the idea.

The secret key $k$, used later by the user to validate the privacy certificate provided by the server, then consists of the $n$ random check bits $s_0,s_1,\dots,s_n$ and their positions. The length of the secret key is, therefore (with the approximation valid when $m\gg n $),
\begin{equation}
\operatorname{len}(k)=	n+\log{{m+n}\choose{n}}\approx n\log{m}\,.
\end{equation}

Note that the encoded message is not, and needs not be, securely encrypted in the traditional sense: An adversary can read almost all of it without requiring the secret key. However, when it does so, it destroys the information needed to conclude successfully the privacy delegation: The eavesdropping can be detected\footnote{This is different from quantum sealing\cite{bechmann2003quantum}, because we require here that the secret key be shorter than the message and do not insist on the message being totally readable by someone not having the key.}. The security parameter $n$ constrains the minimal length $m$ of the message meaningfully stored ($m \gg n$), but it is quite natural to assume the message is long in the storage setting.
\subsection{The Privacy Delegation}
The protocol now splits into two versions according to the chosen task.

\paragraph{Remote-Storage Version}
In the version $\operatorname{STORAGE}$, the server sends back the whole quantum state to the user, who then checks the integrity of the sprinkled bits and separate them from the meaningful data. If the user's measurements on the sprinkled bits match their secret key, they accept the privacy certificate. Otherwise, they reject it and accuse the server of having leaked their information.

Note that the {\em integrity\/} of the information stored on the server is not guaranteed by that basic protocol even when the privacy delegation succeeds because an adversary could for example flip in the rectilinear basis ({\em i.e.}, apply an $\operatorname{X}$-gate) every bit without making $\operatorname{CERT}$ fail. As mentioned before we are, however, not concerned by this kind of attack because it can be prevented in a straightforward way using a Wegman-Carter classical authentication scheme.

\paragraph{Provable-Deletion Version}
In the version $\operatorname{ERASURE}$, the user asks the server to delete all of its data by measuring it in the diagonal basis and publicly announcing the result. If the server behaved honestly and did not leak any data, the output should be completely random except for the sprinkled check bits. The user accepts the privacy certificate if these values correspond to the secret key, and reject it otherwise. They ignore all non-sprinkled bits.

The proof of erasure comes from the server being forced to measure in the diagonal basis bits that are encoded in the rectilinear basis. This gives a series of uniformly random bits while destroying the original information. Since the diagonal-basis measurements are announced publicly by the prover, they cannot rewind the protocol later: The information which was encoded in the rectilinear basis is lost forever.

\subsection{Partial Security}\label{partialsec}

Both versions of the naive protocol offer arbitrarily high partial privacy-delegation security for sufficiently long messages, meaning that a valid certificate information-theoretically guarantees that at most a limited fraction of the total information was leaked. We aim to give the intuition behind this statement by proving it for non-coherent attacks. We show the privacy certificate will be rejected with arbitrarily high probability in presence of enough eavesdropping.
\begin{theorem}
For both $\operatorname{STORAGE}$ and $\operatorname{ERASURE}$, the probability of any prover $\mathcal{P}$ producing a valid privacy certificate in presence of an eavesdropper $\tilde{A}(r)$ doing a rectilinear projective measurement on $r\in\Theta{(m)}$ qubits is given by
\begin{equation}
\operatorname{P}\!\left[\operatorname{CERT}^{(\mathcal{V},\mathcal{P},\mathcal{\tilde{A}}(r))}\right]\le 2^{-rn/(m+n)+\epsilon} + 2\exp(-2\epsilon^2 r) \,,\end{equation}
where $m$ is the message length, $n$ the number of sprinkled check bits, and $\epsilon$ some fixed constant.
\begin{proof}
The probability to pass the certification given the number $K=k$ of randomly sprinkled check bits that were measured by the eavesdropper is $2^{-k}$. $K$ follows a hypergeometric distribution (it is as if the eavesdropper had done classical sampling). The maximal probability of passing the certification is thus
\begin{equation}
\operatorname{P}\!\left[\operatorname{CERT}^{(\mathcal{V},\mathcal{P},\mathcal{\tilde{A}}(r))}\right]=\sum_{k=0}^{r}	2^{-k} \frac {{n\choose k}{m\choose {r-k}} }{ {{m+n \choose r}}}\,.
\end{equation}
	The upper bound follows from Hoeffding's inequality \cite{hoeffding1963,bouman2010sampling},
\begin{equation}
P\left[|K-\mu(K)|\ge \epsilon \right]	\le 2 \exp(-2\epsilon^2 r)\,,
\end{equation}
with $\mu(K)=rn/(m+n)$ since $K$ is hypergeometric.
\end{proof}
\end{theorem}

\newcommand{\VPAB}{(\mathcal{V},\mathcal{\bar{P}},\mathcal{\bar{A}})}
\subsection{An Attack Leaking Partial Information}\label{wattack}
An attack on a limited number of bits is, however, still possible. More precisely, an adversary can pass with non-negligible probability both a discrimination experiment ({\em e.g.}, discriminating a message starting by 0 from an uniformly random one) and the certification experiment. 
\begin{theorem}\label{attack}
$\operatorname{STORAGE}$ and $\operatorname{ERASURE}$ are insecure against an eavesdropper $\mathcal{\bar{A}}$ measuring only, in the rectilinear basis, the first bit of the quantumly encoded message (of length $m$) and a prover $\mathcal{\bar{P}}$ proceeding honestly otherwise.
\begin{proof}
\begin{align*}
\operatorname{P}&\!\left[\operatorname{DISCR}^\mathcal{\VPAB}(n)| \operatorname{CERT}^\mathcal{\VPAB}(n)\right]-1/2\\&\hspace{160pt}\ge (1-n/(n+m))/4 \,,  \\
\operatorname{P}&\!\left[\operatorname{CERT}^\mathcal{\VPAB}(n)\right]\!= 1-n/(2(n+m))\,,\\
\therefore &\operatorname{P}\!\left[\operatorname{CERT}^{\VPAB}\!(n)\right]\nonumber\\& \cdot \left( \operatorname{P}\!\left[\operatorname{DISCR}^{\VPAB}(n)|\operatorname{CERT}^{\VPAB}\!(n)\right]-1/2\right) \not< \operatorname{negl}(n) \,.
\end{align*}
\end{proof}
\end{theorem}

This does not mean that the protocol is useless, as it is already desirable to restrict the amount of stored data that can be leaked without the user becoming aware. This is especially so if the data are already encrypted with computational security. However, for our protocol to be more than only partially secure, the protocol run on any message should be indistinguishable from the one run on any other one. It is important to note the weakness exhibited in Thm.~\ref{attack} is also present in naive implementations of BB84. We will discuss next how this problem is, however, usually resolved in the case of BB84, and what this implies for our privacy-delegation protocol.

\subsection{Privacy Amplification by Public Discussion}\label{privacyampl}
In its original version\cite{BB84}, the BB84 quantum key-distribution protocol was also susceptible to the weak attack suggested in Proposition~\ref{attack}. There was, however, a fix: privacy amplification by public discussion~\cite{pa1,pa2,pa3}. These now well-known schemes aim to reduce the amount of information an eavesdropper can have about the private key. Two parties do so by assessing through public discussion how much information the eavesdropper can have about their key and by agreeing accordingly on a randomly selected error-correcting code or a hashing function from a universal class. When successful (if Eve does not have too much information), the result is a shorter but fully private key. Public discussion between Alice and Bob is achieved through a perfectly authenticated, but not private, channel.

In our storage scenario, public discussion is unavailable as it would necessitate communication between a user's past and future selves. Communication from the past to the future could still be simulated by the user if they keep some secret information, but the amount of this information they would need to privately store to apply standard privacy-amplification schemes would be longer than the message they want to store. Privacy amplification by public discussion as it is usually done fails, therefore, to extend partial privacy-delegation security to full information-theoretic privacy-delegation security in the storage setting. Could there be an alternative way in this setting to amplify the partial security?
And how close to information-theoretic security can we get in presence of more sophisticated adversaries?\footnote{Indistinguishability seems for example to be too strong of a security requirement if the adversary can build coherent attacks using unlimited computational power. That is because the short-key requirement implies that
$\sum_{k\in\mathcal{K}}\rho(M,k)/|\mathcal{K}|$ cannot be uniform over all messages.} These are our main open questions.

\section{Conclusion}
We approached this work with one question in mind: Does quantum mechanics make provable deletion of classical data achievable? This leads us to define rigorously {\em privacy delegation}, an alternative to the ideal of privacy through encryption, and to formalize how to detect leaks during data storage.

We suggest that quantum-provable deletion could be possible by providing a naive privacy-delegation scheme for both the remote-storage and provable-deletion problems. The question remains, however, open as our scheme offers only partial security: It fails against restricted attacks on targeted bits. This shortcoming is known from quantum key distribution; unfortunately, the fix used there, namely privacy amplification by public discussion, cannot be applied in our setting.

To conclude, we emphasize that even if the protocols presented are out of reach of current quantum technologies ---~quantum memory is hard~---, there is real value
in devising and analyzing this kind of theoretical puzzles: Formal definitions give focus and direction to future research, while attempts to solve difficult problems under novel constraints invariably spark new ideas that ultimately turn into the building blocks needed to shrink the domain of the mathematical, cryptological and physical \emph{impossibles}.

\section*{Acknowledgment}
We would like to thank Alberto Montina, Arne Hansen, Bart van der Vecht, Boris \v{S}kori\'c, and Cecilia Boschini for helpful discussions, and the two anonymous reviewers for their useful comments.

\bibliographystyle{IEEEtranbetter}
\bibliography{IEEEabrv,bibthiefknot}

\end{document}